# Cracking predictions of lithium-ion battery electrodes by X-ray computed tomography and modelling

*Adam M. Boyce*[a,b], *Emilio Martínez-Pañeda*[c], *Aaron Wade*[a,b], *Ye Shui Zhang*[a,b,d], *Josh J. Bailey*[a,b,e], *Thomas M. M. Heenan*[a,b], *Dan J. L. Brett*[a,b], *and Paul R. Shearing*[a,b]∗

a. Electrochemical Innovation Lab, Department of Chemical Engineering, University College London, London WC1E 7JE, UK

b. The Faraday Institution, Quad One, Becquerel Ave, Harwell Campus, Didcot, OX11 0RA, UK

c. Department of Civil and Environmental Engineering, Imperial College London, London SW7 2AZ, UK

d. School of Engineering, University of Aberdeen, Aberdeen, AB24 3UE, UK

e. School of Chemistry and Chemical Engineering, Queen's University Belfast, BT7 1NN, UK

*Corresponding author e-mail: p.shearing@ucl.ac.uk

**Abstract**

Fracture of lithium-ion battery electrodes is found to contribute to capacity fade and reduce the lifespan of a battery. Traditional fracture models for batteries are restricted to consideration of a single, idealised particle; here, advanced X-ray computed tomography (CT) imaging, an electro-chemo-mechanical model and a phase field fracture framework are combined to predict the void-driven fracture in the electrode particles of a realistic battery electrode microstructure. An electrode is shown to exhibit a highly heterogeneous electrochemical and fracture response that depends on the particle size and distance from the separator/current collector. The model enables prediction of elevated cracking due to enlarged cycling voltage windows, cracking as a function of electrode thickness, and increasing damage as the rate of discharge is increased.

This framework provides a platform that facilitates a deeper understanding of electrode fracture and enables the design of next-generation electrodes with higher capacities and improved degradation characteristics.

# 1. Introduction

Lithium-ion batteries (LIBs) are at the forefront of the effort to reduce global CO2 emissions. For example, the transition from the fossil fuel-based internal combustion engine to the electrical vehicle (EV) is growing at an increasingly rapid rate, a transition in which LIBs play a pivotal role. The lifespan of any technology is a crucial factor when sustainability is concerned, and for such a pivotal technology as the LIB, it is no exception. The lifespan of LIBs is known to be severely limited by cracking of the electrode components [1,2]. The electrodes within LIBs are a complex multi-material composite and their microstructure typically comprises active particles in which lithium is extracted or inserted, a binder that provides a conductive network for electronic conductivity, and a porous network that permits conduction of lithium-ions via an electrolyte. During charge or discharge, swelling or shrinking of the particles occurs due to insertion or extraction of lithium, giving a highly heterogeneous stress state. Consequently, pre-existing particle defects such as voids or cracks promote fracture, enabling disintegration of the particles and eventual loss of usable energy storage volume [3]. The microstructural evolution of LIB electrodes is therefore central to overall battery performance and, in particular, the long-term degradation and lifespan.

Mechanics plays a significant role in both solid-state lithium transport and the electrochemical reactions occurring at particle surfaces, two processes that are fundamental to LIB operation. The reaction rate at the particle surface and solid-state lithium transport are both influenced by the chemical potential, a component of the system's free energy. The mechanical strain energy contributes to the free energy of the system and thus plays a role in regulating the chemical reactions [4]. The diffusion-related contribution is linked to the hydrostatic stress gradient, which, along with the concentration gradient, provides a driving force for lithium transport within the active material lattice sites. When insertion occurs, we see compressive stresses towards the surface of particles; whilst internally, we see tensile stretching, which generally promotes intercalation [5]. Strain levels may reach ~2-3% in LiCoO2 (LCO) [6] and $LiNiMnCoO_2$ (NMC) [7] based cathodes, 5% in graphite anodes [8], and up to 310% for next-generation Si-based anodes [9]. Irrespective of the electrode material, significant levels of stress develop in an electrode during cycling, and consequently result in the fracture and subsequent degradation of the LIB.

In broad terms, fracture may initiate in three ways: within the primary particles (intragranular), at the grain boundaries between primary particles (intergranular), or due to microstructural defects such as pre-existing voids and cracks in the aggregated secondary particles [10,11]. The three mechanisms are likely to act in tandem. However, in the present study, we focus exclusively on the large voids and cracks that result from synthesis and manufacture; given that they are typically the largest defect feature relative to the primary particle and grain boundaries, they are likely to be the driving force for the majority of damage in an electrode. Fracture of electrode particles contributes to the general degradation of the LIBs through mechanisms such as: build-up of solid-electrolyte-interphase (SEI) on newly cracked surfaces, leading to capacity fade [12]; loss of contact between particles and carbon binder, resulting in electrical isolation of particles, and thus, loss of storage volume; and delamination from the current collector resulting in electrical isolation of the entire electrode [13].

The effort to develop physics-based models and predict the performance of LIBs is a multiscale problem and can be divided into three broad categories: single particle level, micro-scale electrode level, and cell level. The single particle [14-16] and cell [17-19] models typically employ homogenised versions of the physical equations used in three-dimensional micro-scale models [20] due to their computational efficiency and accuracy. These models have since been extended to consider electro-chemo-mechanical coupling and stress-induced diffusion [21,22], with further additions to account for capacity fade due to factors including SEI growth [23]. Image-based micro-scale modelling tends to consider electrode microstructure heterogeneity, and comparatively less work has been carried out at this level with electro-chemo models [24,25] and then mechanics-based contributions [26-28].

Numerical predictions of crack growth typically treat the crack as a discrete discontinuity in a material and are implemented using techniques such as cohesive zone elements where the crack advances along a predefined path [27, 29-31]. Methods such as X-FEM can predict arbitrary crack trajectories [32,33] but are limited when dealing with complex cracking conditions, particularly in 3D. In contrast, the continuum-based phase field approach does not require prior knowledge of crack location and can predict complex conditions of practical applications, which may include arbitrary crack trajectories, non-sharp defects, merging of voids, cracks and other defects, etc. For example, phase field models have been developed for prediction of fracture in hydrogen embrittled steel for deep-sea oil pipelines [34,35], shape memory alloys for medical stents [36], as well as nuclear reactor materials [37], amongst numerous others. The phase field approach has also been used to simulate fracture in idealised LIB single particles [37-43]. Other computational methods, such as X-FEM [32,33], cohesive elements [27, 29-31] or continuum damage mechanics [44] have also been used to predict the fracture of single particles. However, the ability of phase field to capture complex cracking patterns in arbitrary geometries and dimensions make it particularly well-suited to achieve a step-change in LIB simulation and predict the behaviour at the electrode level, including the interactions resulting from multiple components (particles, pores, binders). Xu et al. [27] carried out electro-chemo-mechanical simulations of an image-based NMC electrode but the fracture analysis was limited to a single (idealised) particle. Here, for the first time, we propose combining the phase field framework with advanced imaging techniques and the aforementioned electro-chemo-mechanical models to deliver realistic multiphysics predictions of damage and degradation of an entire electrode. Specifically, these techniques are combined in order to predict the evolution of damage within a realistic representation of a LIB electrode ($LiNi_{0.6}Mn_{0.2}Co_{0.2}O_2$, NMC622).

The model represents a step-change from the traditional model of a single particle in which we capture the coupled electrochemical, mechanical and damage performance of an entire electrode. This is achieved using a combination of advanced 3D X-ray CT imaging and an electro-chemo-mechanical formulation that includes the phase field framework for fracture predictions. We model the coupled electro-chemo-mechanical behaviour and fracture of the active secondary particles, the mechanical behaviour of the PVDF-carbon-binder composite domain (CBD), and their interactions with the current collector. The simulations undertaken here advance the fundamental understanding of battery fracture and its dependency on electrode morphology, particle position and electrochemical cycling. This knowledge enables the design of next-generation electrodes with higher capacities and improved degradation characteristics.

## 2. Theoretical framework

Consider the schematic of a battery half-cell in Fig. 1a: we define a separator, electrode surface, current collector and a heterogeneous electrode domain comprising of active material, bound together by carbon additives, i.e. CBD, and micropores filled with electrolyte. Fig. 1b shows the image-based 3D domain to which the following theoretical framework is applied.

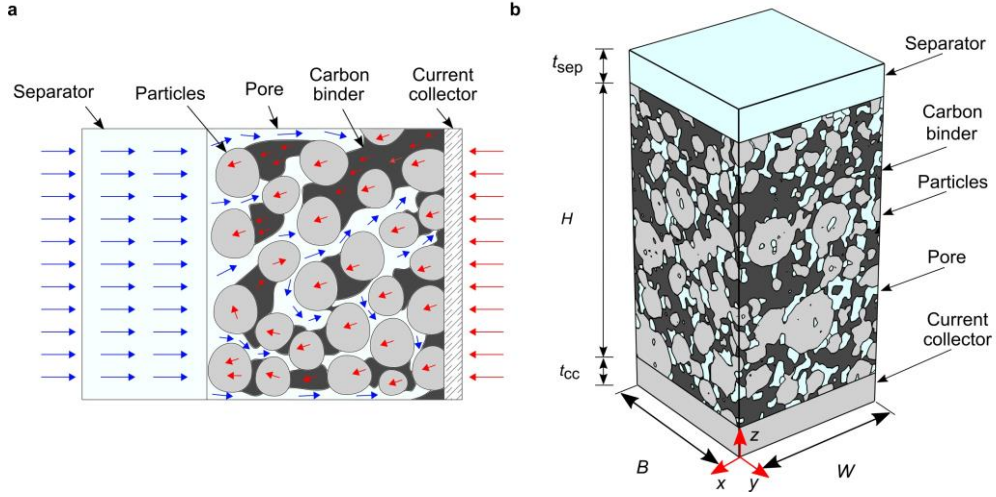

**Fig. 1.** Schematic and image-based representations of LIBs. a) 2D schematic of an electrode microstructure highlighting flow of electrons (red) and flow of lithium ions (blue). b) 3D image-based microstructure captured via X-ray CT, where $B=W=50$ µm, $H=89$ µm, and $t_{cc}=t_{sep}=10$ µm.

*2.1 Electrolyte: electrochemical and transport formulation*

Consider a binary electrolyte described by concentrated solution theory [45]. To form an accurate description of mass transfer within such a medium, we need to consider diffusive mass transfer based on Fick's first law, and migration of ions due to an electric field and the presence of an electrochemical potential field, as described by Thomas-Alyea and Newman [45]. If a 1:1 electrolyte is considered, then a balance of positive lithium ions (concentration, $c_{e+}$) and counterions (concentration, $c_{e-}$) may be assumed, allowing us to write $c_{e+}=c_{e-}=c_e$. The resulting flux of positive ions $\boldsymbol{J}_e$ is given as

$$\boldsymbol{J}_e = -D_e \nabla c_e + \frac{\boldsymbol{i}_e t_+}{F} \tag{1}$$

where $D_e$ is the electrolyte diffusion coefficient, $\boldsymbol{i}_e$ is the ionic current, $t_+$ is the cation transference number (the proportion of current transported via positive lithium ions), and $F$=96485 C mol$^{-1}$, is the Faraday constant. Fick's second law must be invoked to capture the non-steady-state transport of lithium ions

$$\frac{\partial c_e}{\partial t} + \nabla \cdot \boldsymbol{J}_e = 0 \tag{2}$$

The ionic current flow within the electrolyte, as derived based on the Stefan-Maxwell equations by [45], may then be defined as

$$\boldsymbol{i}_e = (-K_e \nabla \phi_e) + \frac{2K_e RT}{F}\left(1 + \frac{\partial \ln f}{\partial \ln c_e}\right)(1-t_+)\nabla \ln c_e \tag{3}$$

This is formed of two components. The first term is the classic description of Ohm's law where $K_e$ is the ionic conductivity and $\phi_e$ is the electric potential across the medium. The second term is based on the electrochemical activity of the ions [45]. $R$ and $T$ are the universal gas constant, and temperature, respectively, whilst $f$ is the activity coefficient that depends on the electrochemical potential. $f$, $t_+$ and $K_e$ are typically given as a function of $c_e$. Finally, the entire electrochemical system, and thus the electrolyte, must observe charge conservation

$$\nabla \cdot \boldsymbol{i}_e = 0 \tag{4}$$

*2.2 Active material: electro-chemo-mechanical and transport formulation*

We consider the stress field in the active particles to be dictated by mechanical equilibrium, given as follows in the absence of body forces

$$\nabla \cdot \boldsymbol{\sigma} = 0 \tag{5}$$

where $\boldsymbol{\sigma}$ is the Cauchy stress tensor. For NMC particles, such as those considered in this study, it is appropriate to apply a simple linear elastic material model [27]. Thus the elastic strain $\boldsymbol{\varepsilon}^e$ is governed by Hooke's law; $\boldsymbol{\sigma}=\mathbf{C}:\boldsymbol{\varepsilon}$. $\mathbf{C}$ is the elastic stiffness matrix, and in this instance we consider an isotropic model with Young's modulus, $E_p$ and Poisson's ratio, $v$=0.3. Applying a small strain formulation, the total strain $\boldsymbol{\varepsilon}$ in an active particle is obtained by solving for the displacement field $\boldsymbol{u}$ in the solid

$$\boldsymbol{\varepsilon} = \frac{1}{2}\left((\nabla\boldsymbol{u})^T + \nabla\boldsymbol{u}\right) \tag{6}$$

It can be recognised that the total strain may be decomposed into the elastic strain $\boldsymbol{\varepsilon}^e$ and the chemical, lithiation-induced strain $\boldsymbol{\varepsilon}^{ch}$ as follows

$$\boldsymbol{\varepsilon} = \boldsymbol{\varepsilon}^e + \boldsymbol{\varepsilon}^{ch} \tag{7}$$

The lithiation-induced strain is

$$\boldsymbol{\varepsilon}^{ch} = \frac{1}{3}\Omega\left(c_p - c_{p0}\right)\mathbf{I} \tag{8}$$

where $\Omega$ is the partial molar volume of the active material, $c_{s0}$ is the concentration of Li in the unstressed state, and $\mathbf{I}$ is the identity matrix that recognises that the contribution from lithiation-induced strains produces a hydrostatic stress state only.

*2.2.1 Phase field fracture formulation*

Consider an arbitrary domain $B$ where a crack exists with corresponding cracked surfaces $\Gamma$. The variation of the total energy $\Pi$ due to an incremental increase in the crack area $dA$ is a function of the elastic strain energy density of the solid $\psi(\boldsymbol{\varepsilon}^e)$ and the energy required to create two new surfaces $W_s$, as dictated by thermodynamics. In the absence of external forces:

$$\frac{d\Pi}{dA} = \frac{d\psi}{dA} + \frac{dW_s}{dA} = 0 \tag{9}$$

where $G_c = dW_s/dA$ is the critical energy release rate that characterises the resistance to fracture of a material. It follows that we can recast Equation (9) in variational form as follows [46]

$$\Pi = \int_\Omega \psi(\boldsymbol{\varepsilon}^e) dV + \int_\Gamma G_c d\Gamma \qquad (10)$$

To overcome the fact that the crack surface is unknown, we now introduce a phase field variable $\phi$ that assumes the value of 1 for a fully cracked material point and 0 when the material is uncracked. Furthermore, a damage degradation function $g$ can be defined that enables the diminishing stiffness of the material to be quantified, where $g(\phi) = (1-\phi)^2$. Griffith's functional can then be approximated as [47]

$$\Pi = \int_\Omega \left[ (1-\phi)^2 \psi(\varepsilon^e) + G_c \left( \frac{1}{2l}\phi^2 + \frac{l}{2}|\nabla\phi|^2 \right) \right] dV \qquad (11)$$

where $l$ is the phase field length scale, which characterises the size of the fracture process zone and provides a relation with the material strength [48]. By forming the Lagrangian $L(u, \dot{u}, \phi)$ and finding a stationary point $\delta L = 0$, the following balance equation for the phase field can be found

$$G_c \left( \frac{1}{l}\phi - l\frac{\partial\phi}{\partial x^2} \right) - 2(1-\phi)\psi(\varepsilon^e) = 0 \qquad (12)$$

Equation (12) constitutes the evolution equation for the phase field, which enables the prediction of cracking behaviour of arbitrary complexity as a result of the exchange between elastic and fracture energies.

*2.2.2 Solid-state transport*

The diffusive flux of lithium ions in the particles is governed by Fick's first law of diffusion, which is modified to account for the driving force presented by the presence of hydrostatic stresses

$$\boldsymbol{J}_p = -D_p \left( \nabla c_p + \frac{\Omega c_p}{RT} \nabla \sigma_H \right) \qquad (13)$$

where $\sigma_H = \text{tr}[\boldsymbol{\sigma}]/3$. Transient transport of the lithium within the solid particles is given by Fick's second law

$$\frac{\partial c_p}{\partial t} + \nabla \cdot \boldsymbol{J}_p = 0 \qquad (14)$$

The particles follow the typical charge conservation law

$$\nabla \cdot \boldsymbol{i}_p = 0 \qquad (15)$$

whilst electronic conduction is governed by Ohm's law,

$$\boldsymbol{i}_p = -K_p \nabla \phi_p \qquad (16)$$

*2.2.3 Charge transfer reaction*

Here we impose a boundary condition at the interface between particle and electrolyte to capture the charge transfer reaction. Once positive lithium ions reach the surface of the particles, they are neutralised by electrons during a charge transfer reaction and subsequently diffuse into the particles. The charge transfer rate can be modelled by a Butler-Volmer type equation

$$i_{BV} = i_0 \left( \exp\left(\frac{\alpha_a F \eta}{RT}\right) - \exp\left(-\frac{\alpha_c F \eta}{RT}\right) \right) \qquad (17)$$

The local exchange current density is defined as a function of $c_p$ and $c_e$

$$i_0 = F(k_c)^{\alpha_a}(k_a)^{\alpha_c}(c_p - c_{p,\max})^{\alpha_a}(c_p)^{\alpha_c}\left(\frac{c_e}{c_{e,\text{ref}}}\right)^{\alpha_a} \qquad (18)$$

where $c_{e,\text{ref}}$ is a reference electrolyte concentration, typically taken as unity. The deviation from equilibrium conditions is governed by

$$\eta = \phi_p - \phi_e - U - \frac{\Omega \sigma_H}{F} \qquad (19)$$

where $\eta$ is the overpotential, $U$ is the open circuit voltage, and the fourth term on the right-hand side accounts for the diffusion-induced hydrostatic stress on the overpotential.

## 2.3 Carbon binder domain: electrical, transport and mechanical formulation

Finally, we consider the conductive carbon binder phase: the binder maintains electrical contact between the particles and the current collector permitting electron flow, thereby facilitating the charge transfer reaction. In reality, the carbon binder domain phase is a nanoporous foam-like material, for example see the work of Daemi et al. [49] where the carbon binder was measured to be ~28% porous. To minimise computational expense, we attribute effective transport properties, which permits both ionic (lithium ions) and electronic flow. The conductive phase must obey Ohm's law

$$\boldsymbol{i}_c = -K_c \nabla \phi_c \tag{20}$$

where $\boldsymbol{i}_c$ and $\phi_c$ are the current and potential in the CBD, respectively. Charge must also be conserved; therefore

$$\nabla \cdot \boldsymbol{i}_c = 0 \tag{21}$$

Ionic transport within the electrolyte has been outlined in Section 2.1. The effective conductive properties of carbon binder $K_{c,\text{eff}}$, effective ionic conductivity $K_{e,\text{eff}}$ of electrolyte and effective diffusivity $D_{e,\text{eff}}$ of the electrolyte within the nanopores, are determined by scaling their associated bulk properties by a factor *f=0.276*. The factor *f* is equivalent to the porosity factor of typical carbon binder material as reported by Daemi et al. [49].

The mechanical behaviour of the CBD is assumed to be isotropic, undergo small strains and adhere to a simple linear elastic Hooke's law [26] with Young's modulus $E_c$ and Poisson's ratio, *v*=0.3. The strains are obtained via Equation (6) with no lithiation-induced strains present. An ideal boundary, i.e. perfect contact, is assumed between the active materials and CBD. A similar ideal boundary is assumed between the CBD and the current collector which is given as an electronically conductive, linear elastic solid undergoing small strains.

## 2.4 Boundary and initial conditions

Consider Fig. 1b; we define a coordinate system ($x$, $y$, $z$) as shown. At the current collector, we apply a current density $\boldsymbol{i}_c \cdot \boldsymbol{n}_{cc} = i_{in}$ at $z=0$ where $\boldsymbol{n}_{cc}$ is the unit normal vector pointing outwards from the current collector surface in the negative $x$-direction. The applied current density $i_{in}$ at 1C discharge, is calculated based on the active material volume $V_p$, the maximum accepted quantity of lithium for the given electrode material, $c_{p,max}$: the current density is given as $i_{in}= c_{p,max}FV_p/t_0 A$, where $A=BW$ and $t_0=3600$ s. In addition, we prescribe fully clamped conditions to the current collector surface, i.e. $\boldsymbol{u} \cdot \boldsymbol{n}_{cc} = 0$ whilst all other surfaces are free.

A flux of electrolyte salts is applied at the separator: $\boldsymbol{J}_1 \cdot \boldsymbol{n}_{el} = i_{in}/F$ at $z=t_{cc}+L+t_{sep}$, where $\boldsymbol{n}_{el}$ is the unit normal vector pointing from the separator in the positive $z$-direction. $t_{cc}$, $L$, $t_{sep}$ are the thicknesses of the current collector, electrode and separator, respectively. In addition, an ionic current is applied at the separator: $\boldsymbol{i}_1 \cdot \boldsymbol{n}_{el} = -i_{in}$ at $z=t_{cc}+L+t_{sep}$. A potential of $\phi_p=0$ is applied at the same location. At the interface between the separator and heterogeneous electrode we specify that the electronic current flow must be zero; $\boldsymbol{i}_c \cdot \boldsymbol{n}_{se} = 0$ and $\boldsymbol{i}_p \cdot \boldsymbol{n}_{se} = 0$ at $z=L+t_{cc}$ where $\boldsymbol{n}_{se}$ is the unit normal vector to the interface between the electrode and the separator, pointing in the positive $z$-direction. This ensures that only the ionic current is permitted across this surface.

At the electrolyte-particle interface we observe a flux of lithium into the solid, or lithium ions into the electrolyte as a result of the charge transfer reaction. The fluxes are as follows: $\boldsymbol{J}_e \cdot \boldsymbol{n}_{pe} = -i_{BV}/F$ and $\boldsymbol{J}_s \cdot \boldsymbol{n}_{pe} = -i_{BV}/F$, where $\boldsymbol{n}_{pe}$ is the normal vector pointing from the electrolyte to the active particles. We also prescribe a current density at this interface: $\boldsymbol{i}_e \cdot \boldsymbol{n}_{pe} = -i_{BV}$ and $\boldsymbol{i}_p \cdot \boldsymbol{n}_{pe} = i_{BV}$.

An initial active material concentration, $c_{p0}$, is prescribed, whilst the electrolyte initial concentration is given by $c_{e0}$. The electrode and all associated constituent domains are assumed to be in an initially unstressed state. Finally, all external domain surfaces normal to the *x-z* plane and *y-z* plane are assumed to be insulating to active material- , CBD- , and electrolyte-related current densities, as well as species fluxes.

## 3. Experimental methods

### 3.1 Electrode manufacture

The formulations for the cathode made in this work are generic formulations that are extensively employed [50,51]. The cathode slurry consists of $LiNi_{0.6}Mn_{0.2}Co_{0.2}O_2$ (NMC622, BASF), polyvinylidene fluoride (PVDF, Solvay) and C65 (Imerys) powders, which were pre-dried at 120 °C in a vacuum oven over 12 h to remove any moisture. 13 mL of N-methyl-2-pyrrolidone (NMP, Sigma Aldrich) mixed with 0.4 g of PVDF was then weighed to form a binder solution. A THINKY mixer (ARE-20, Intertronics) was used to mix the cathode binder solution (PVDF and NMP) at 2000 rpm for 15 min until the solution became homogeneous. Then, 19.3 g of NMC622 and 0.4 g of C65 were added slowly to the binder solution to form a slurry with a solid content ~60 wt. %. The slurry was then mixed again with THINKY mixer at 2000 rpm for 2 periods of 15 min. The slurry was left to cool between the two mixing steps to avoid any temperature increment of the slurry. Finally, the homogeneous slurry was degassed in the THINKY mixer at a speed of 2000 rpm for 2 min.

The cathode slurry was then coated on a piece of aluminium foil with thickness ~16 µm (PI-KEM) using a doctor blade thin-film applicator (calibrated with a metal shim), resulting in a dried electrode of ~89 µm thickness.

### 3.2 Image acquisition

The NMC622 sample was prepared using an A Series/Compact Laser Micromachining System (Oxford Lasers, Oxford, UK) with an embedded Class 4, 532 nm wavelength laser [52]. X-ray CT imaging was conducted using a lab-based X-ray micro-CT instrument (ZEISS Xradia 520 Versa, Carl Zeiss., CA, USA) with a tube voltage of 80 kV. An exposure time of 28 s was used for each of the 601 projections, with a 20× magnification lens. Reconstruction of the data was carried out via Zeiss XM software (Carl Zeiss., CA, USA), utilizing cone-beam filtered back-projection algorithms, resulting in a voxel size of 0.371 µm.

## 4. Numerical methods

*4.1 Image processing*

The raw CT image was transformed, cropped and filtered (using a non-local means filter) within Avizo software (Avizo, Thermo Fisher Scientific, Waltham, Massachusetts, USA). In order to balance the microstructural resolution and computational expense, we chose a region of interest with the full electrode thickness $L$=89 µm and $B$=$W$=50 µm. A single slice from the processed image can be seen in Fig. 2a. Machine learning based, open-source segmentation software ilastik (Berg et al. [53]) was used to segment the three-phase raw image (pore, CBD and particles). The segmentation process yielded a ternary electrode image with the following volume fractions: 51.3% particle, 35.5% carbon binder, and 13.2% pore. The segmentation results are shown in Fig. 2b (the same slice as Fig. 2a). We have used established image analysis protocols to extract the relative volume fractions of the active material, CBD and macro-porosity, and have achieved vol% which are comparable to literature benchmarks [49]. Regions representing the current collector and separator were added to the segmented volume with $t_{cc}$=$t_{sep}$=10 µm.

*4.2 Finite element implementation*

Simpleware ScaniP (Mountain View, CA, USA) was used to mesh the segmented image (see Fig. 2c), giving approximately 3.6 million linear tetrahedral elements, with 11.6 million degrees of freedom. The characteristic element size in the regions of interest is 400 nm, which is 4.5 smaller than the phase field length scale and thus sufficient to achieve mesh convergence [34,54]. The meshed volume is shown in Fig. 2c. The theoretical framework was implemented in the finite element software COMSOL Multiphysics (v5.6, Sweden) on a 3D tomography-based mesh, as described. The Parallel Direct Sparse Solver (PARDISO) was used to solve the discretised transport, electrode kinetics and deformation kinematics equations. A segregated approach was taken, which involved solving the coupled field variables in a sequential staggered manner, see Miehe et al. for details [54]. Also, a history field was defined to prohibit damage irreversibility [54]. Time stepping was handled using $2^{nd}$ order backward Euler differentiation, whilst time step sensitivity analysis was carried out.

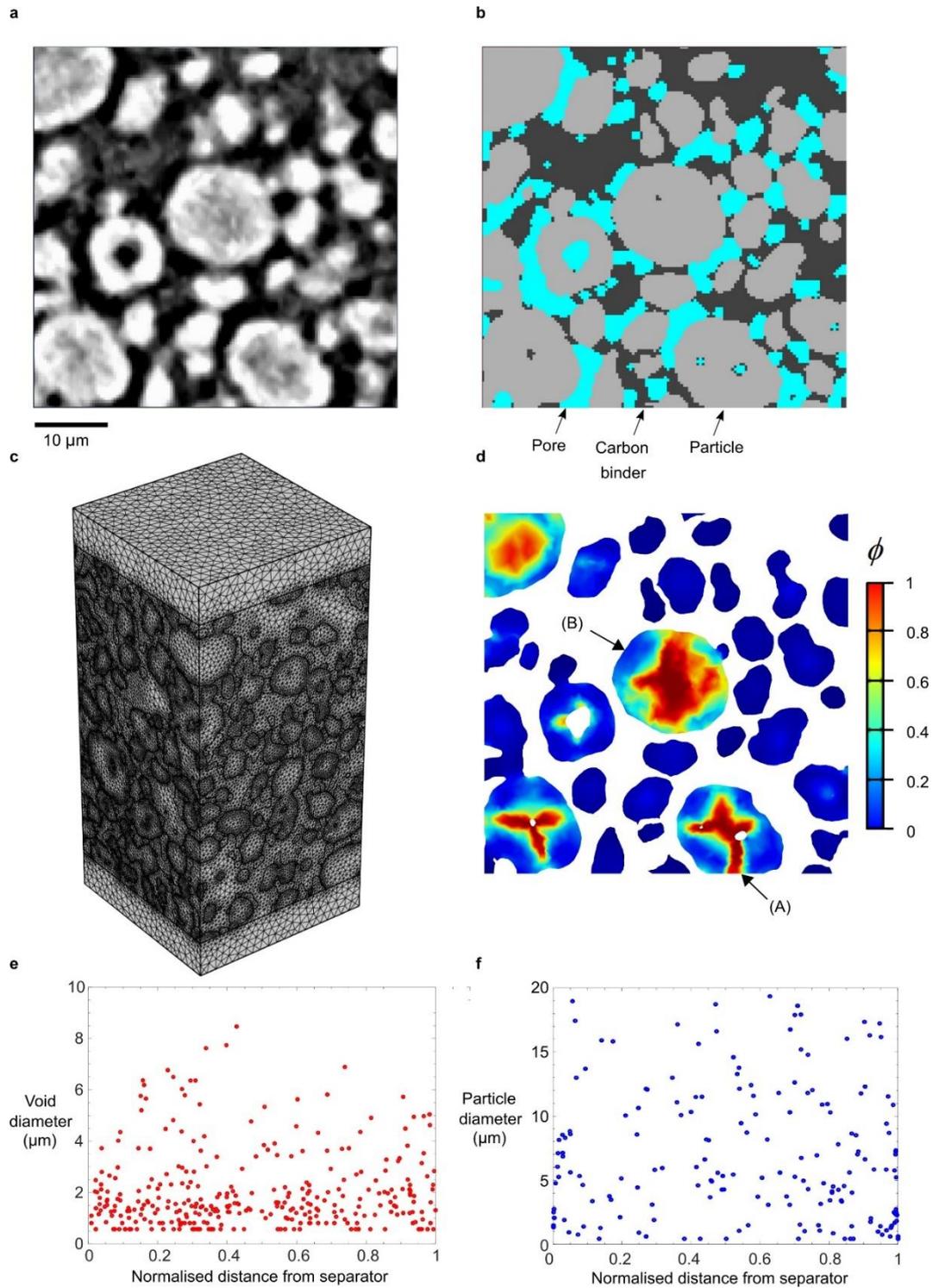

**Fig. 2.** X-ray tomography, segmentation, meshing and simulation of an electrode. a) Tomogram of cropped three-phase image, b) segmentation of CT image, c) computational mesh of segmented 3D volume, d) example of model output showing fracture at a slice equivalent to a) and b), e) void size, and f) particle size distributions as a function of electrode thickness.

*4.3 Material models*

Properties for the constituent materials were extracted from literature and are outlined in **Table 1**. NMC622 was treated as an isotropic linear-elastic solid with an underlying assumption of brittle failure. The Young's modulus and fracture toughness of NMC were obtained from [27] while we estimated the phase field length parameter, *l*, through its relation with the material strength $\sigma_c$, as described by Tanné et al. [48]. The solid-state diffusivity of NMC622 was treated as non-linear, with a dependence on lithium concentration, drawn from the measurements of Noh et al. [55]. The open-circuit voltage response of NMC622 was drawn from the work of Xu et al. [27]. The carbon binder was assumed to be an electronically conductive, linear elastic material with a Young's modulus magnitude following the measurements by Grillet et. al. [56]. A $LiPF_6$ electrolyte was modelled and the salt concentration-dependent properties were given by Valoen and Reimers [57]. The aluminium current collector was assumed to be an electronically conductive linear elastic solid with electrical conductivity of $K=3.7\times10^7$ S m$^{-1}$, Young's modulus of $E$=69 GPa, and Poisson's ratio of $\nu$=0.3. Finally, the universal gas constant was given as 8.314 J mol$^{-1}$ K$^{-1}$, whilst all simulations were carried out assuming a temperature of $T$=293 K.

**Table 1.** Material properties and model parameters.

| Parameter | Unit | Value | Source |
|---|---|---|---|
| $t_+$ | 1 | 0.37 | Valoen et al. [57] |
| $\dfrac{\partial \ln f}{\partial \ln c_e}$ | 1 | 0.43 | Valoen et al. [57] |
| $D_e$ | m$^2$ s$^{-1}$ | f($c_e$) | Valoen et al. [57] |
| $K_e$ | S m$^{-1}$ | f($c_e$) | Valoen et al. [57] |
| $c_{e0}$ | mol m$^{-3}$ | 1000 | -- |
| $D_p$ | m$^2$ s$^{-1}$ | f($c_p/c_{pmax}$) | Noh et al. [55] |
| $c_{p,max}$ | mol m$^{-3}$ | 48700 | Xu et al. [27] |
| $K_p$ | S m$^{-1}$ | 1.6×10$^{-4}$ | Noh et al. [55] |
| $E_p$ | GPa | 140 | Xu et al. [27] |
| $G_c$ | N m$^{-1}$ | 0.11 | Xu et al. [27] |
| $\Omega$ | m$^3$ mol$^{-1}$ | 1.8×10$^{-6}$ | Koerver et al. [58] |
| $c_{p0}$ | mol m$^{-3}$ | 500 | -- |
| $V_p$ | m$^3$ | 1.3×10$^{-13}$ | -- |
| $\sigma_c$ | MPa | 100MPa | Xu et al. [27] |
| $l$ | m | 1.8×10$^{-6}$ | Tanné et al. [48] |
| $K_c$ | S m$^{-1}$ | 375 | Liu et al. [59] |
| $E_c$ | GPa | 0.3 | Grillet et al. [56] |
| $\alpha_a$ | 1 | 0.5 | -- |
| $\alpha_b$ | 1 | 0.5 | -- |
| $k_a$ | m s$^{-1}$ | 2×10$^{-11}$ | Xu et al. [27] |
| $k_b$ | m s$^{-1}$ | 2×10$^{-11}$ | Xu et al. [27] |
| $U$ | V | f($c_p/c_{pmax}$) | Xu et al. [27] |

*4.4 Simulation details*

We applied the described model to a variety of test cases that allowed us to gain insight into the electrode fracture behaviour. The three cases were as follows:

1. A single high rate discharge at 6C, followed by other single discharges at 1C, 3C and 9C to establish the influence of rate of discharge on electrode fracture. All discharges began at the pristine state with no prior cycling history.
2. The influence of electrode thickness on fracture behaviour (at 6C discharge rate).
3. The influence of voltage window and multi-cycle charges and discharges on the extent of electrode cracking (at 6C discharge rate).

## 5. Results and discussion

An example of the phase field fracture profile following high-rate discharge (lithiation of the positive electrode) at 6C is shown in Fig. 2d (on the same slice as Fig. 2a and b) - a systematic study of C-rate follows. With the increasing need for fast charging of EVs and potential extreme operating conditions, e.g. severe accelerations, it is prudent to assess high discharge (and charge) rates such as those assessed in this study. Furthermore, there is the possibility of local heterogeneities in discharge rate within the microstructure, potentially resulting in excessive cracking and degradation. At a discharge rate of 6C, even in the presence of small central voids, Fig. 2d reveals that the model predicts significant fracture within an electrode as widely reported [30,60]. Our hypothesis of a void-driven fracture process is backed up by fracture mechanics estimates of the critical flaw size. For a material strength of $\sigma_c$=100 MPa, Young's modulus $E_p$=140 GPa and energy release rate of $G_c$=0.11 N/m (as outlined in Table 1), then the transition flaw size $a_t$ can be calculated as $a_t = K_c / \pi \sigma_c^2$ =0.49 μm, where $K_c = (EG_c)^{0.5}$ =0.124 MPa m$^{0.5}$. As shown in Fig. 2e, a large number of voids have a characteristic length larger than this transition flaw size. Moreover, we observe that the extent of the cracks in any given particle (average measured particle radius ~4.8 µm) is quite large and that particles will completely fracture with cracks extending from central voids to the outer surfaces of the particles (see particle A in Fig. 2d), whilst large disk-like cracks will form due to central cracks (see particle B in Fig. 2d). These observations qualitatively agree with those reported by Klinsmann et al. [10] in their analysis of sharp cracks in perfectly spherical single particles. Later, we will comment on this in the context of a non-idealised sphere, i.e. the as-manufactured particle.

As a precursor to the ensuing studies we must first highlight the electrochemical behaviour of the electrode, see Fig. 3a. It is clear that a significant reduction in usable capacity occurs when the electrode is discharged at increasingly higher rates. This is a widely reported issue, see for example, Park et al. [61]; here, the influence of the electrode thickness is also investigated. In the case of a 6C discharge we observe that an 89 µm thick electrode has 38% less usable capacity when compared to the 44.5 µm thick electrode. Reductions in lithium storage capacity due to rate behaviour as a function of electrode thickness have been reported by Xu et al. [62].

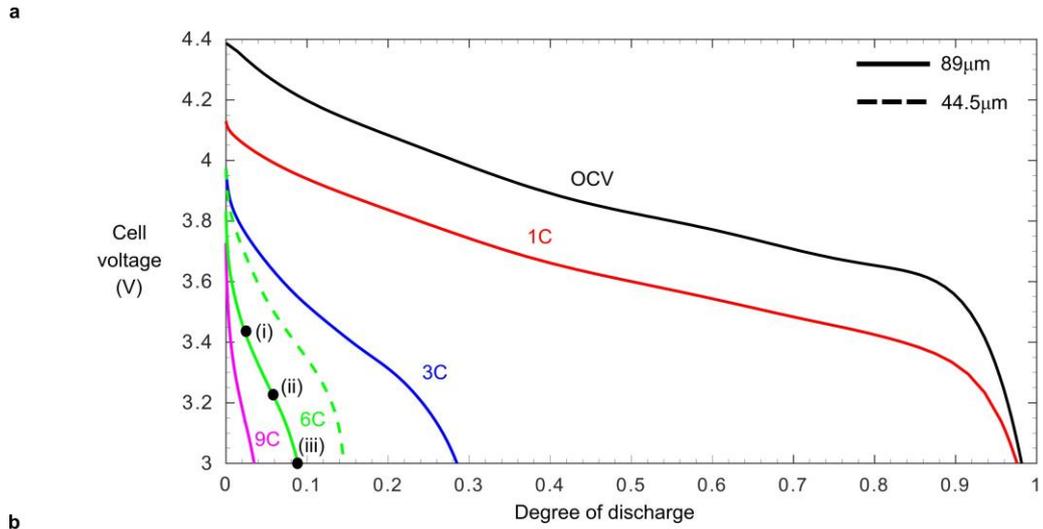
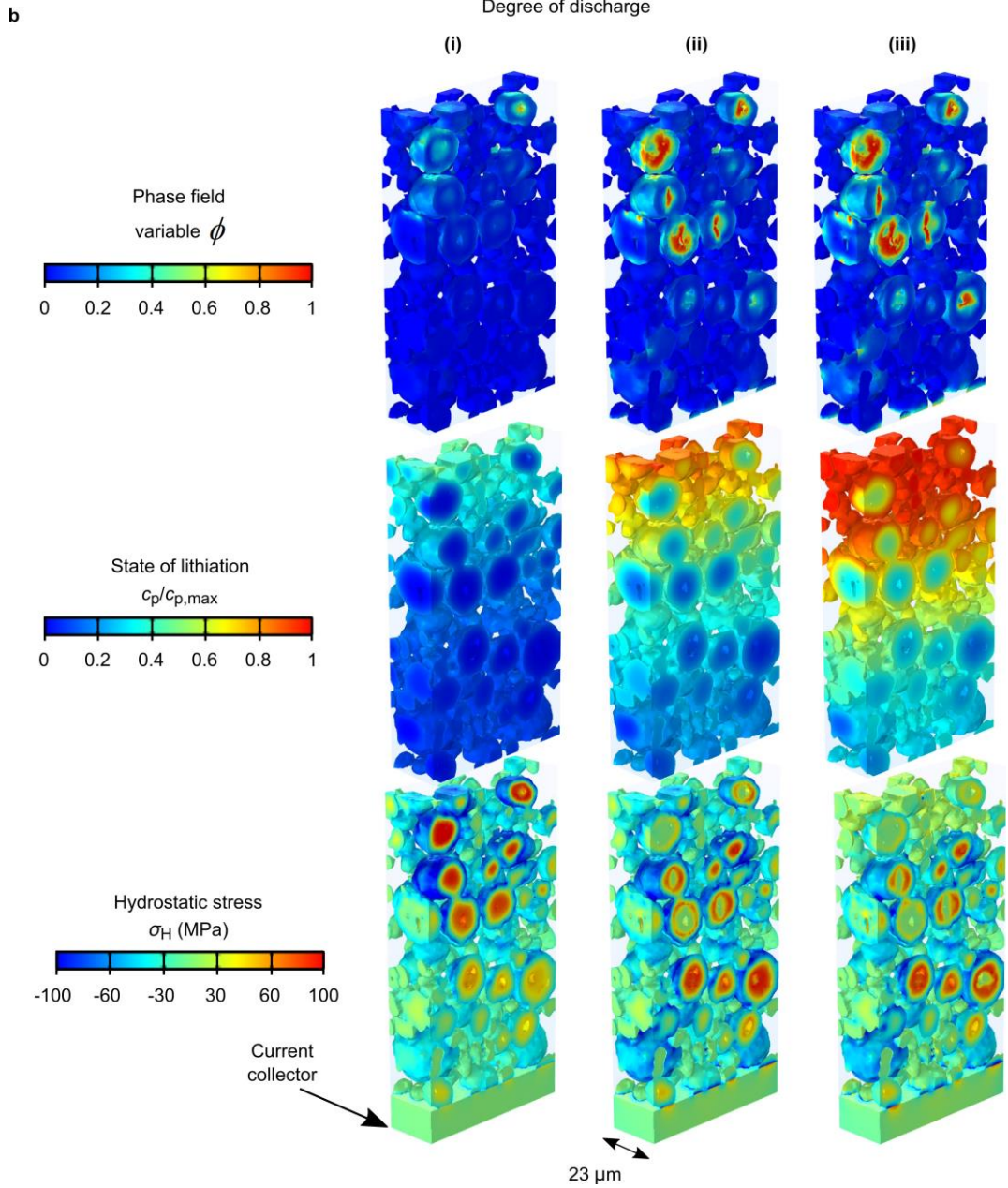

**Fig. 3.** Simulated electrochemical and fracture response of a LIB. a) The open circuit voltage and the simulated voltage response as a function of degree of discharge for the electrode volume at 1C, 3C, 6C, and 9C. b) Phase field, lithium concentration and hydrostatic stress distributions in a sample volume at three instances of discharge as highlighted in a) i), ii) and iii).

## 5.1 General observations: first discharge

Fig. 3a illustrates the predicted electrode voltage response during a single discharge, i.e. lithiation of NMC, at a rate of 6C with a voltage window of 3.0 V-3.8 V. We highlight three key regions on the voltage response. First, at ~20% degree of discharge, labelled as (i), we observe crack initiation in some large particles (radii~6-7 µm) close to the separator (highlighted in Fig. 3b(i)). This corresponds with severe lithium concentration and hydrostatic stress gradients throughout the electrode. Furthermore, we observe lithium concentration and hydrostatic stress gradients within the particles themselves (primarily within the particles close to the separator). This behaviour is linked to diffusion-limited transport. At high discharge rates, the electrode is experiencing significant lithium ion flux in the near-separator region. However, this high concentration of lithium ions cannot be distributed homogeneously throughout the electrode due to the limitations of the electrolyte diffusivity. A similar rationale explains the concentration gradients within the particles themselves: once lithium ions are neutralised at the surface of particles, lithium must diffuse to the centre of the particles where there is a low concentration of lithium relative to the surface. If a particle begins in a delithiated state then there will be a significantly higher lithium concentration near the surface of the particle due to diffusive transport limitations. This imposes tensile hydrostatic stress towards the centre of the particle as a result of the excessive lithium in the outer regions of the particle. This heterogeneous concentration and stress state may be observed in Fig. 3b (i). It is clear that particle size dictates the severity of concentration and stress gradients. Thus, elevated tensile stresses in the centre of the particles induces fracture that is further exacerbated by the presence of voids, which act as stress raisers. Similar qualitative behaviour was observed for 2D individual idealised particles [63].

As discharge proceeds, we observe progressive fracture of the electrode, note points (ii) and (iii) in Fig. 3a and the associated phase field profiles in Fig. 3b. Particle-level concentration and stress gradients move as a front through the material throughout discharge, resulting in fracture of the inner regions of the electrode. At point (iii), the end of discharge, the regions closest to the current collector remain in a delithiated state relative to the regions in the centre and closer to the separator. This under-utilisation of active material results from transport limitations and the depletion of lithium ions in the electrode due to the high surface current densities in the separator regions during early discharge. Consequently, the level of particle damage is lowest in the current collector region and at its peak closest to the separator; this has been observed experimentally by Xu et al. [27] and Heenan et al. [60].

*5.2 Influence of discharge rate*

The rate of discharge influences the permissible energy stored [61]. Furthermore, it is generally accepted that an increased rate of charge or discharge, i.e. rate of lithiation/delithiation, reduces the lifespan of the battery [50]. Discharge or charge of cells at a high rate increases the extent of concentration gradients within the particles, resulting in higher levels of particle stresses. We illustrate this in Fig. 4, which highlights the post-discharge fracture patterns in the electrode at four different rates. At 1C rate of discharge with a 3.0 V-4.3 V voltage window, we observe no obvious particle cracking. The cracking is primarily restricted to the narrow regions or struts that connect the particles. In the case of 9C discharge, we observe severe particle cracking both at the narrow strut regions and within the particles themselves. In the intermediate cases, we observe an increased level of cracking, which may be attributed to the increasing degree of heterogeneity in the lithium concentration and thus the hydrostatic stress profiles. Furthermore, we observe that as the rate of discharge increases the fracture is increasingly limited to the separator region.

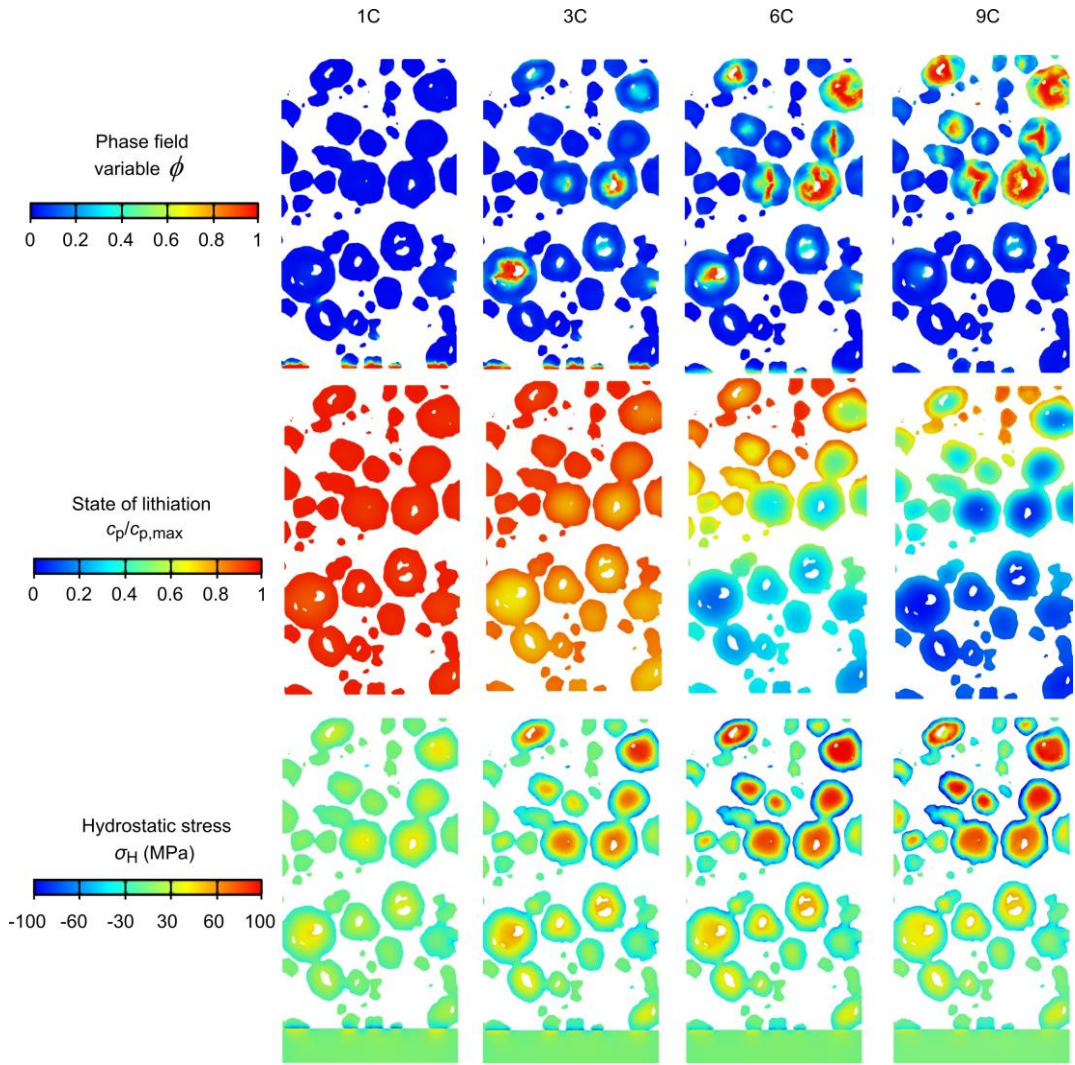

**Fig. 4.** The influence of discharge rate on electrochemical and fracture performance. A slice through the electrode volume with profiles of the phase field variable, lithium concentration at the end of discharge at rates of 1C, 3C, 6C, and 9C. The hydrostatic stress profile is shown at approximately 10% degree of discharge and prior to fracture initiation.

*5.3 Influence of electrode thickness*

Subsequently, as shown in Fig. 5, we alter the thickness of the electrode by cropping the original image (using Avizo) to observe the influence of a shorter diffusion length on fracture behaviour.

The particle size distribution through the thickness of the original 89 µm electrode is shown in Fig. 2f. It appears that the particle size is randomly distributed throughout the thickness of the electrode, with no clear gradient. Furthermore, given the uniform distribution of flaws within the electrode (Fig. 2e), cropping of the image is likely to have no significant influence on the crack initiation. The thinner electrode, discharged at 6C, displays severe cracking throughout the entire microstructure. There are no longer through-thickness, electrode-level gradients in lithium concentration although the particle-level gradients exist throughout the electrode, causing fracture. It is clear that the lithium concentration gradients throughout the thin electrode are less pronounced, relative to the thicker electrode, which is commensurate with the lower capacity of the thicker electrode when discharged at high rate. In addition, we note that the hydrostatic stress gradient, while detrimental in terms of fracture, acts as a driving force for lithium transport within the particles. It follows that a more uniform distribution in large hydrostatic stresses at the electrode-level will facilitate better utilisation of the entire electrode volume.

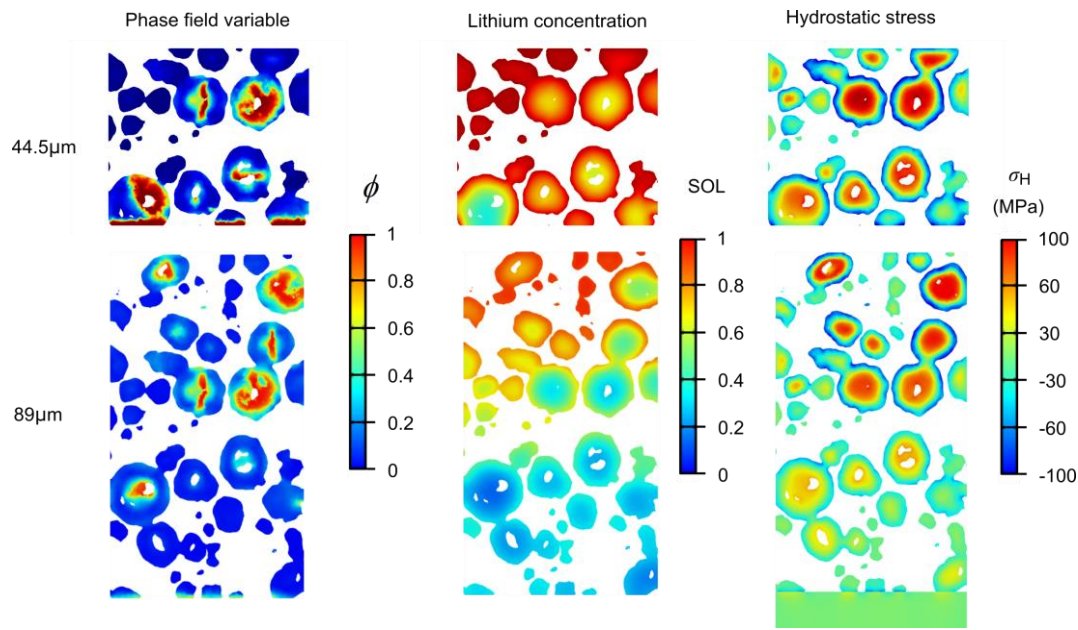

**Fig. 5.** The influence of electrode thickness on electrochemical and fracture performance. A slice through two electrodes of 44.5 µm and 89 µm thickness, with profiles of the phase field

variable, lithium concentration following discharge and hydrostatic stress prior to fracture initiation at approximately 10% depth of discharge. The discharge rate was 6C.

*5.4 Influence of cycling and depth of charge/discharge*

We now consider the microstructural evolution when cycling of the electrode is simulated. Consider first the electrochemical response (I) in Fig. 6a. The electrode was charged and discharged at a rate of 6C between 3.0 V and 4.3 V over 5 cycles. First, we observe that a severe loss in capacity occurs after the first cycle. This may be attributed to significant polarisation, which is represented by a sharp increase in voltage, and happens at the beginning of the first charge (and then at the beginning and end of each subsequent half cycle). This effectively reduces the cycling voltage window and results in reduced opportunity to remove or insert lithium or lithium ions to or from the particles. Here we show that once the first high-rate discharge is complete, a subsequent charging results in a limited level of delithiation or removal of lithium from the particles (see the lithium concentration profiles of Fig 6b, (I), End of $1^{st}$ charge). We note that, throughout cycling, the polarisation is greater during charge (cathode delithiation). This asymmetry in polarisation leads to an observed asymmetry in the duration of charge and discharge. Thus, more lithium is being added to the particles than extracted due to the diminished opportunity for electrochemical reactions. Furthermore, we observe capacity fade such that the duration of both charge and discharge half cycles decreases progressively as cycling proceeds. This can be interpreted via profiles of active material lithium concentration at the end of the $5^{th}$ cycle (Fig 6b, (I), End of $5^{th}$ discharge). There is a greater degree of homogeneity in lithium concentration throughout the electrode, giving a lower gradient and thus driving force for diffusion of lithium within the particles, resulting in the aforementioned capacity fade. Experimental evidence [64] has shown that the overpotential is greater during charge and that this contributes to capacity fade due to increasing inefficiency of lithium ion transport or potential slowing down of the charge transfer. This aligns with the mechanism of capacity fade presented here.

We now compare the above results with those of a larger voltage window (3.8 V-4.5 V), see Fig. 6a (II) and the corresponding lithium concentration profiles in Fig. 6b (II). Here, the larger voltage window permits greater levels of lithiation and delithiation, resulting in greater utilisation and storage capacity. However, we note that there is still a lower capacity upon continuous cycling of the electrode, albeit to a lesser extent than the lower voltage window (Case I). Thus the same polarisation-related mechanism dominates here.

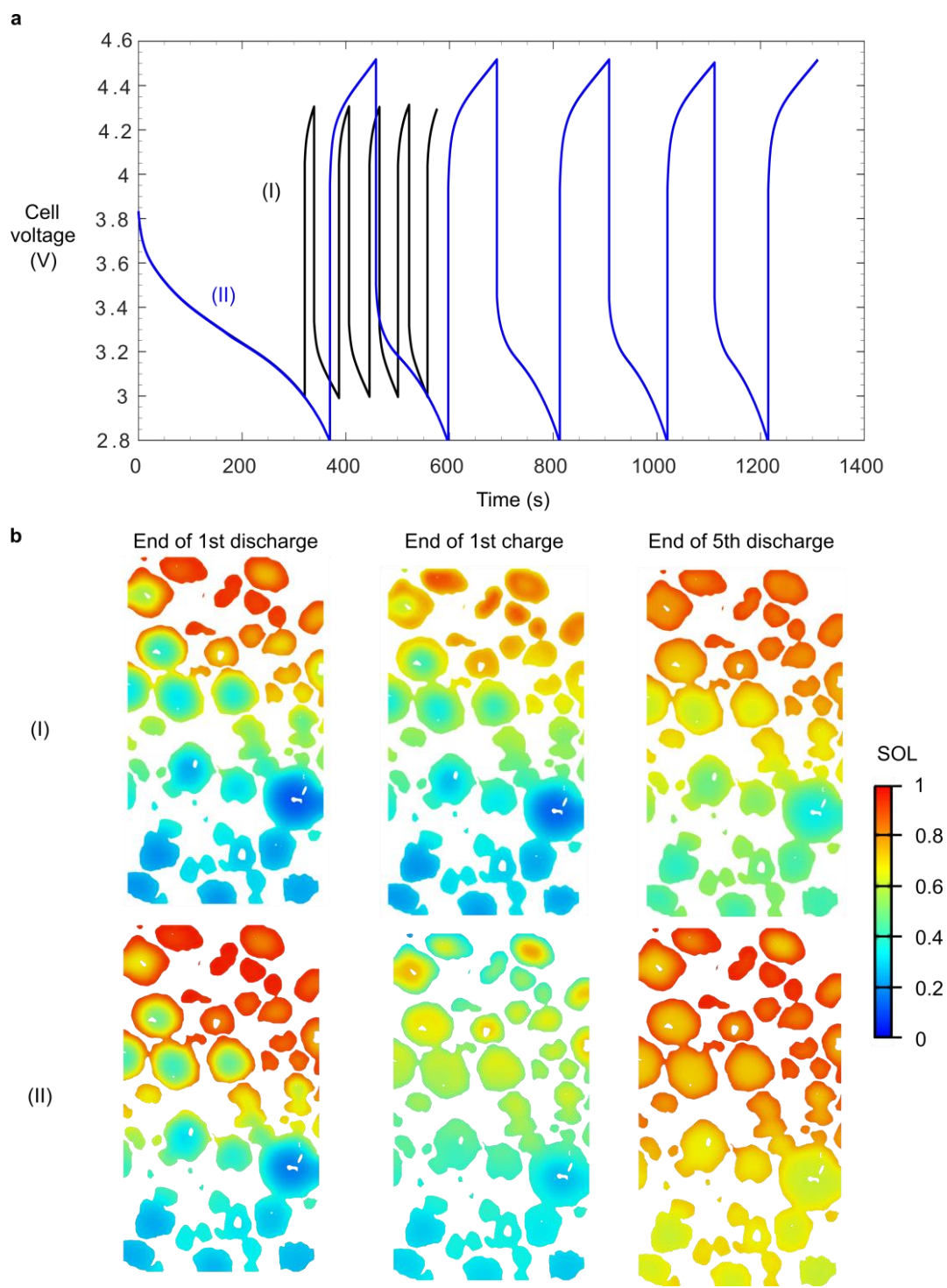

**Fig. 6.** The electrochemical response during cycling of an electrode. **a** Voltage response of an 89 µm thick electrode undergoing cycling with voltage window of Case (I) 3 V-4.3 V and Case (II) 2.8 V-4.5 V. **b** The lithium concentration profiles at the end of the 1st discharge, end of 1st charge and end of 5th discharge for Cases (I) and (II).

Fig. 7 shows the evolution of fracture as cycling proceeds. It is clear in both cases (I) and (II) that the majority of damage occurs within the first cycle. With a greater depth of discharge (discharge to 2.8 V in Case (II)), we observe that a larger degree of cracking occurs relative to that of Case (I), the discharge to 3 V. Upon charging and then further cycling we observe additional crack growth, most of which occurs in the $1^{st}$ charge, see particles (X) and (Y) in Fig. 7, for example. The results from above show that a larger voltage window facilitates greater extents of lithiation and delithiation, giving greater capacity. However, there are drawbacks to this, as we observe greater degrees of damage in the centre of the electrode upon cycling, see particle (Y) in Fig. 7 (II), for example.

During delithiation, we observe a reversal in particle-level hydrostatic stress gradients such that the outer surfaces of the particles are in tension and the central regions are in compression. In this scenario, the presence of edge defects will promote crack growth [42]. In the present study, the majority of defects are located in the centre of the particles and since we begin from an undamaged, fully delithiated (charged) state, the majority of fracture occurs there. Any cracks that develop during discharge will extend further during charge due to the rearrangement of the stress field, and this is the primary mechanism presented here.

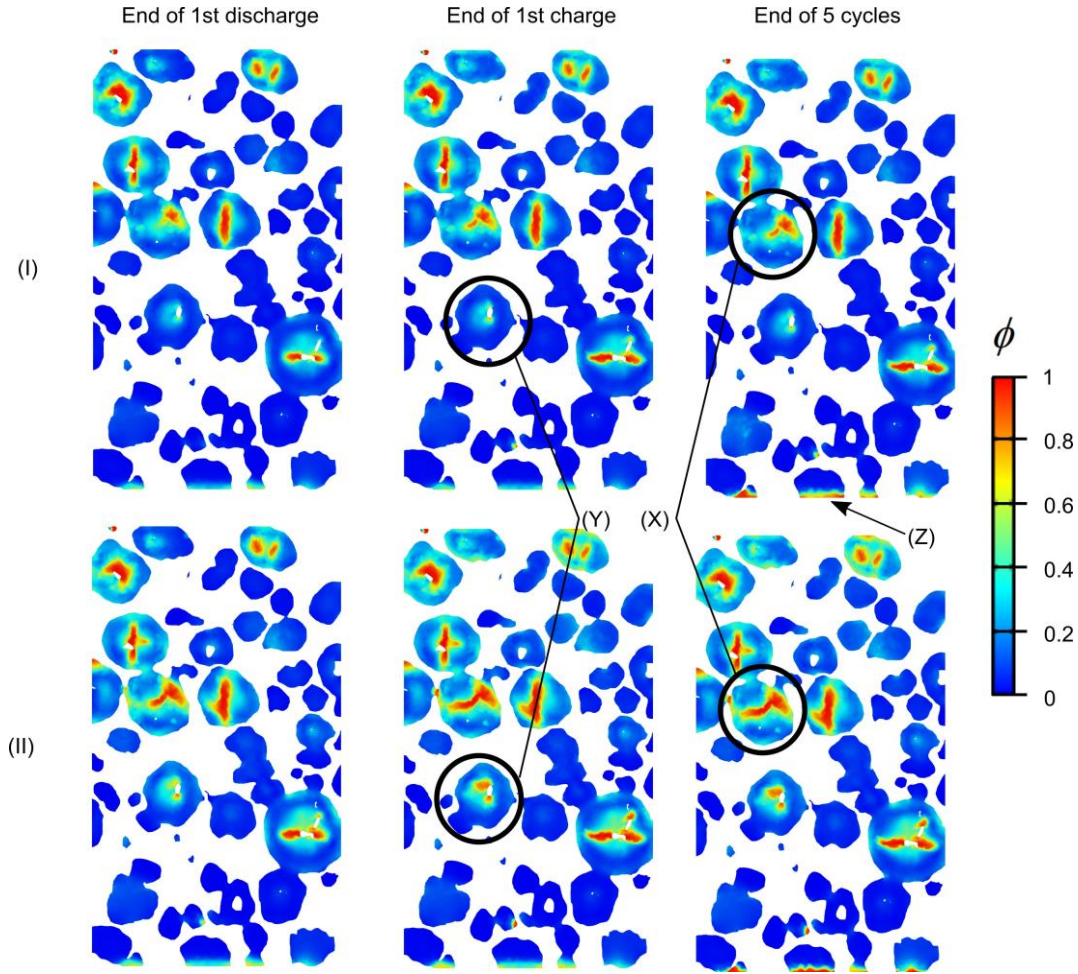

**Fig. 7.** The fracture response during cycling of an electrode. The phase field profile at the end of the 1st discharge, end of 1st charge and end of 5th cycle for Cases (I) and (II) in Fig. 6.

*5.5 Damage at the current collector-particle interface*

A common observation and theme within all simulations was the fracture response at the interface between the aluminium current collector and the attached NMC particles, see for example point (Z) in Fig. 7. It is clear that significant levels of tensile stress develop at the interface due to expansion of the particles. In addition, the attached particles are, by nature, not completely spherical, and this results in stress concentrations at the boundary between the current collector, particle and the surrounding CBD or pore. Some caveats must be kept in mind as we proceed with this discussion. We do not model the post-yield stress-strain behaviour of aluminium, nor do we consider the bond strength between the NMC particles and aluminium (we assume they are perfectly attached, i.e. they share the same computational nodes at the point of contact). However, we can determine that there is the potential for damage within the NMC particles at this interface and furthermore the potential for detachment from the current collector, which results in electrical isolation of the electrode. This behaviour has been observed to occur during the cycling of NMC electrodes by Heenan et al. [60]. Cracking at the current collector interface exhibits a dependence on discharge rate; Fig. 4 shows that cracking at this interface is more likely to occur at lower C-rates (1C and 3C for example). The absence of interface cracking at higher discharge rates can be attributed to the greater degree of heterogeneity in electrode-level lithium concentration and hydrostatic stresses, as described previously. However, we note that once an electrode is cycled at high rate, this interface cracking occurs at later cycles (see Fig. 7) due to the ever-increasing lithium concentration in particles adjacent to the current collector as cycling proceeds.

*5.6 Observations on individual particle fracture*

Shown in Fig. 8 is the detailed mechanical and fracture characteristics of a representative single particle (radius~6.9 µm) that has been extracted from the full electrode simulations, as described above. As the discharge at 6C progresses, we observe that fracture progresses in three stages, as follows: (1) cracks initiate at pre-existing voids and do so due to the high levels of triaxial stress, as indicated by the coexistence of highly tensile first, second and third principal stresses in the centre of the particle. (2) Following crack initiation, we observe unstable crack growth and coalescence of the cracks into a single central disk-like void (see Fig. 8(ii) and (iii)), also observed by Klinsmann et al. [10] for an ideal insertion particle with multiple voids. (3) Growth of the central void continues and multiple instances of branching occur. The cracks then arrest without fully reaching the particle surface. We observe that the stress is still compressive at the outer regions of the particle, which prevents further crack growth. During delithiation upon charging, the outer surfaces will exhibit tensile hydrostatic stresses, which increases the likelihood of further propagation of these cracks towards the surface.

The branching in (3) may be interpreted in the following manner. When a crack occurs there is stress relief at the crack surfaces, which results in redistribution of the stress field in the vicinity of the crack faces. We can observe this via the changes that occur in the principle, and thus hydrostatic stresses, as shown in Fig. 8(ii) and (iii). However, it is evident that there are sufficient tensile stresses adjacent to the crack tip to promote further crack growth. Given that fracture occurs perpendicular to the direction of maximum principle stress, this further crack growth and its associated branching is therefore due to the change in direction of the maximum principle stress. Furthermore, we observe that crack branching occurs in arbitrary directions, which may be partly attributed to the asymmetric nature of the particle geometry, the heterogeneous distribution of carbon binder at the particle surface, and the resulting asymmetric lithium concentration distributions that we observe in Fig. 8. The influence of carbon binder is not explored in detail in this study, however some assumptions can be made on this subject: the volume fraction of CBD and Young's modulus are likely to influence the overall microstructural stiffness, which will, in turn, affect the level of particle expansion/contraction during lithiation/delithiation. Consequently, this may influence the level of damage experienced by the microstructure, however, this is beyond the scope of the present study and further work is required to ascertain the influence of these factors on particle fracture. In broad terms, the fracture behaviour described above is indicative of all particles throughout the electrode, some of which are shown in Fig. 3.

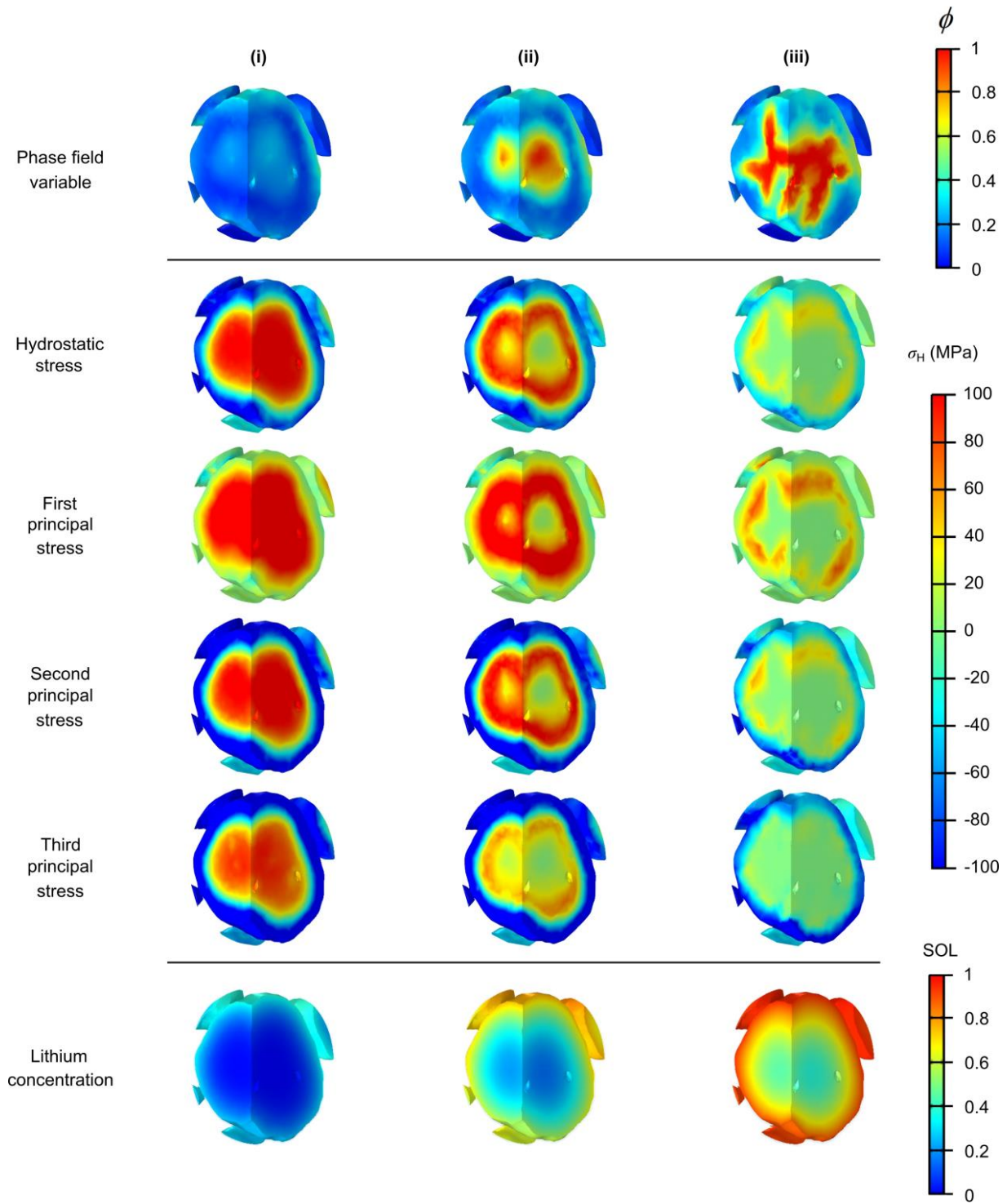

**Fig. 8.** The electrochemical, mechanical and fracture response of a single particle. A particle (radius~6.9 µm) extracted from the electrode volume during discharge at 6C, where (i)-(iii) are the same discharge instances as that in Fig. 3. Profiles of phase field, hydrostatic and principal stresses, and lithium concentration are shown.

# 6. Conclusion

The coupled electro-chemo-mechanical and phase field formulation presented in this work has demonstrated the ability to predict the void-driven damage that occurs within the realistic microstructure of a lithium-ion battery during early-stage cycling. Moreover, we demonstrate the ability of the model to predict the microstructural evolution at a variety of cycling rates and electrode thicknesses, whilst also considering the cycling voltage window. Incidence of damage is shown to be highly location-dependent within the electrode microstructure, which, is linked to the heterogeneous electrochemical response due to slow diffusion in the electrolyte and active material. Furthermore, the majority of void-induced damage is shown to occur within the first cycle, with relaxation of the structure thereafter.

With these electrochemical, transport and mechanical insights in mind, it is now possible to propose some potential electrode design and operating guidelines. For example, given that slow diffusion within the particles limits the performance of thick electrodes and promotes damage, then it is clear that, in particular, smaller particles are required in the near-separator region. This suggests that grading of the electrode microstructure may be required to produce high capacity, thicker electrodes, with improved degradation characteristics. In addition, it is evident that the voltage window must be limited, given the propensity for cracking when an electrode is operated at the extremities of the voltage window during charge and discharge.

We have demonstrated that a framework formed of advanced imaging and coupled numerical implementations of electrochemistry, mechanics and phase field fracture provides a wealth of information about the characteristics of a lithium-ion battery. This paper has outlined the foundations for a tool that will benefit the design of improved battery electrode microstructures and will help define operational boundary conditions to preserve battery performance.

**Acknowledgements**

This work was carried out with funding from the Faraday Institution [EP/S003053/1, grant number FIRG001 and FIRG015]. PRS would like to acknowledge the Royal Academy of Engineering [CiET1718\59] for financial support.

**CRediT author contribution statement**

**Adam M. Boyce:** Conceptualization, Methodology, Software, Investigation, Data Curation, Formal analysis, Writing-Original Draft. **Emilio Martínez-Pañeda:** Conceptualization, Software, Writing - Review & Editing. **Aaron Wade:** Methodology, Investigation, Writing - Review & Editing. **Ye Shui Zhang:** Methodology, Investigation, Writing - Review & Editing **Josh J. Bailey:** Investigation, Writing - Review & Editing. **Thomas M. M. Heenan:** Investigation, Writing - Review & Editing. **Dan J. L. Brett:** Supervision, Project administration, Funding acquisition, Writing - Review & Editing. **Paul R. Shearing:** Conceptualization, Supervision, Project administration, Funding acquisition, Writing - Review & Editing.

**Declaration of competing interest**

The authors declare no competing financial interests.

**References**

[1] C.R. Birkl, M.R. Roberts, E. McTurk, P.G. Bruce, D.A. Howey, Degradation diagnostics for lithium ion cells, J. Power Sources. 341 (2017) 373–386. https://doi.org/10.1016/j.jpowsour.2016.12.011.

[2] P. Li, Y. Zhao, Y. Shen, S.H. Bo, Fracture behavior in battery materials, JPhys Energy. 2 (2020) 022002. https://doi.org/10.1088/2515-7655/ab83e1.

[3] Y. Yang, R. Xu, K. Zhang, S.J. Lee, L. Mu, P. Liu, C.K. Waters, S. Spence, Z. Xu, C.

Wei, D.J. Kautz, Q. Yuan, Y. Dong, Y.S. Yu, X. Xiao, H.K. Lee, P. Pianetta, P. Cloetens, J.S. Lee, K. Zhao, F. Lin, Y. Liu, Quantification of Heterogeneous Degradation in Li-Ion Batteries, Adv. Energy Mater. 9 (2019) 1900674. https://doi.org/10.1002/aenm.201900674.

[4]  L. Anand, A Cahn-Hilliard-type theory for species diffusion coupled with large elastic-plastic deformations, J. Mech. Phys. Solids. 60 (2012) 1983–2002. https://doi.org/10.1016/j.jmps.2012.08.001.

[5]  Y. Zhao, P. Stein, Y. Bai, M. Al-Siraj, Y. Yang, B.X. Xu, A review on modeling of electro-chemo-mechanics in lithium-ion batteries, J. Power Sources. 413 (2019) 259–283. https://doi.org/10.1016/j.jpowsour.2018.12.011.

[6]  H. Mendoza, S.A. Roberts, V.E. Brunini, A.M. Grillet, Mechanical and Electrochemical Response of a LiCoO2 Cathode using Reconstructed Microstructures, Electrochim. Acta. 190 (2016) 1–15. https://doi.org/10.1016/j.electacta.2015.12.224.

[7]  X. Cheng, M. Pecht, In situ stress measurement techniques on li-ion battery electrodes: A review, Energies. 10 (2017) 1–19. https://doi.org/10.3390/en10050591.

[8]  Y. Qi, S.J. Harris, In Situ Observation of Strains during Lithiation of a Graphite Electrode, J. Electrochem. Soc. 157 (2010) A741-A747. https://doi.org/10.1149/1.3377130.

[9]  L.Y. Beaulieu, K.W. Eberman, R.L. Turner, L.J. Krause, J.R. Dahna, Colossal reversible volume changes in lithium alloys, Electrochem. Solid-State Lett. 4 (2001) A137-A140. https://doi.org/10.1149/1.1388178.

[10] M. Klinsmann, D. Rosato, M. Kamlah, R.M. McMeeking, Modeling crack growth during Li insertion in storage particles using a fracture phase field approach, J. Mech.

Phys. Solids. 92 (2016) 313–344. https://doi.org/10.1016/j.jmps.2016.04.004.

[11] P. Barai, P.P. Mukherjee, Stochastic Analysis of Diffusion Induced Damage in Lithium-Ion Battery Electrodes, J. Electrochem. Soc. 160 (2013) A955–A967. https://doi.org/10.1149/2.132306jes.

[12] R. Deshpande, M. Verbrugge, Y.-T. Cheng, J. Wang, P. Liu, Battery Cycle Life Prediction with Coupled Chemical Degradation and Fatigue Mechanics, J. Electrochem. Soc. 159 (2012) A1730–A1738. https://doi.org/10.1149/2.049210jes.

[13] G. Kermani, E. Sahraei, Review: Characterization and modeling of the mechanical properties of lithium-ion batteries, Energies. 10 (2017) 1730. https://doi.org/10.3390/en10111730.

[14] A.M. Bizeray, J.H. Kim, S.R. Duncan, D.A. Howey, Identifiability and parameter estimation of the single particle lithium-ion battery model, IEEE Transactions on Control Systems Technology 27 2019 1862-1877 https://doi.org/10.1109/TCST.2018.2838097.

[15] S.G. Marquis, V. Sulzer, R. Timms, C.P. Please, S.J. Chapman, An asymptotic derivation of a single particle model with electrolyte, J. Electrochem. Soc. 166 (2019) A3693-A3706. https://doi.org/10.1149/2.0341915jes.

[16] D. Zhang, B.N. Popov, R.E. White, Modeling Lithium Intercalation of a Single Spinel Particle under Potentiodynamic Control, J. Electrochem. Soc. 147 (2000) 831-838. https://doi.org/10.1149/1.1393279.

[17] M. Ecker, S. Käbitz, I. Laresgoiti, D.U. Sauer, Parameterization of a Physico-Chemical Model of a Lithium-Ion Battery, J. Electrochem. Soc. 162 (2015) A1849–A1857. https://doi.org/10.1149/2.0541509jes.

[18] M.J. Hunt, F. Brosa Planella, F. Theil, W.D. Widanage, Derivation of an effective thermal electrochemical model for porous electrode batteries using asymptotic homogenisation, J. Eng. Math. 122 (2020) 31–57. https://doi.org/10.1007/s10665-020-10045-8.

[19] V. Srinivasan, C.Y. Wang, Analysis of Electrochemical and Thermal Behavior of Li-Ion Cells, J. Electrochem. Soc. 150 (2003) A98-A106. https://doi.org/10.1149/1.1526512.

[20] M. Doyle, T.F. Fuller, J. Newman, Modeling of Galvanostatic Charge and Discharge of the Lithium/Polymer/Insertion Cell, J. Electrochem. Soc. 140 (1993) 1526–1533. https://doi.org/10.1149/1.2221597.

[21] J. Christensen, J. Newman, Stress generation and fracture in lithium insertion materials, J. Solid State Electrochem. 10 (2006) 293–319. https://doi.org/10.1007/s10008-006-0095-1.

[22] X. Zhang, W. Shyy, A. Marie Sastry, Numerical Simulation of Intercalation-Induced Stress in Li-Ion Battery Electrode Particles, J. Electrochem. Soc. 154 (2007) A910-A916. https://doi.org/10.1149/1.2759840.

[23] J. Li, K. Adewuyi, N. Lotfi, R.G. Landers, J. Park, A single particle model with chemical/mechanical degradation physics for lithium ion battery State of Health (SOH) estimation, Appl. Energy. 212 (2018) 1178–1190. https://doi.org/10.1016/j.apenergy.2018.01.011.

[24] X. Lu, A. Bertei, D.P. Finegan, C. Tan, S.R. Daemi, J.S. Weaving, K.B. O'Regan, T.M.M. Heenan, G. Hinds, E. Kendrick, D.J.L. Brett, P.R. Shearing, 3D microstructure design of lithium-ion battery electrodes assisted by X-ray nano-computed tomography and modelling, Nat. Commun. 11 (2020) 1–13. https://doi.org/10.1038/s41467-020-

15811-x.

[25] X. Lu, S.R. Daemi, A. Bertei, M.D.R. Kok, K.B. O'Regan, L. Rasha, J. Park, G. Hinds, E. Kendrick, D.J.L. Brett, P.R. Shearing, Microstructural Evolution of Battery Electrodes During Calendering, Joule. 4 (2020) 2746–2768. https://doi.org/10.1016/j.joule.2020.10.010.

[26] M.E. Ferraro, B.L. Trembacki, V.E. Brunini, D.R. Noble, S.A. Roberts, Electrode Mesoscale as a Collection of Particles: Coupled Electrochemical and Mechanical Analysis of NMC Cathodes, J. Electrochem. Soc. 167 (2020) 013543. https://doi.org/10.1149/1945-7111/ab632b.

[27] R. Xu, Y. Yang, F. Yin, P. Liu, P. Cloetens, Y. Liu, F. Lin, K. Zhao, Heterogeneous damage in Li-ion batteries: Experimental analysis and theoretical modeling, J. Mech. Phys. Solids. 129 (2019) 160–183. https://doi.org/10.1016/j.jmps.2019.05.003.

[28] P. Huang, Z. Guo, Li-ion distribution and diffusion-induced stress calculations of particles using an image-based finite element method, Mech. Mater. 157 (2021) 103843. https://doi.org/10.1016/j.mechmat.2021.103843.

[29] G. Bucci, T. Swamy, Y.M. Chiang, W.C. Carter, Modeling of internal mechanical failure of all-solid-state batteries during electrochemical cycling, and implications for battery design, J. Mater. Chem. A. 5 (2017) 19422–19430. https://doi.org/10.1039/c7ta03199h.

[30] R. Xu, K. Zhao, Corrosive fracture of electrodes in Li-ion batteries, J. Mech. Phys. Solids. 121 (2018) 258–280. https://doi.org/10.1016/j.jmps.2018.07.021.

[31] S. Lee, J. Yang, W. Lu, Debonding at the interface between active particles and PVDF binder in Li-ion batteries, Extrem. Mech. Lett. 6 (2016) 37–44.

https://doi.org/10.1016/j.eml.2015.11.005.

[32] X. Zhu, Y. Chen, H. Chen, W. Luan, The diffusion induced stress and cracking behaviour of primary particle for Li-ion battery electrode, Int. J. Mech. Sci. 178 (2020) 105608. https://doi.org/10.1016/j.ijmecsci.2020.105608.

[33] M. Zhu, J. Park, A.M. Sastry, Fracture Analysis of the Cathode in Li-Ion Batteries: A Simulation Study, J. Electrochem. Soc. 159 (2012) A492–A498. https://doi.org/10.1149/2.045204jes.

[34] E. Martínez-Pañeda, A. Golahmar, C.F. Niordson, A phase field formulation for hydrogen assisted cracking, Comput. Methods Appl. Mech. Eng. 342 (2018) 742–761. https://doi.org/10.1016/j.cma.2018.07.021.

[35] P.K. Kristensen, C.F. Niordson, E. Martínez-Pañeda, A phase field model for elastic-gradient-plastic solids undergoing hydrogen embrittlement, J. Mech. Phys. Solids. 143 (2020) 104093. https://doi.org/10.1016/j.jmps.2020.104093.

[36] M. Simoes, E. Martínez-Pañeda, Phase field modelling of fracture and fatigue in Shape Memory Alloys, Comput. Methods Appl. Mech. Eng. 373 (2021) 113504. https://doi.org/10.1016/j.cma.2020.113504.

[37] W. Li, K. Shirvan, Multiphysics phase-field modeling of quasi-static cracking in urania ceramic nuclear fuel, Ceram. Int. 47 (2021) 793–810. https://doi.org/10.1016/j.ceramint.2020.08.191.

[38] C. Miehe, H. Dal, L.-M. Schanzel, A. Raina, A phase-field model for chemo-mechanical induced fracture in lithium-ion battery electrode particles, Int. J. Numer. Methods Eng. (2016) 683–711. https://doi.org/10.1002/nme.5133.

[39]  A. Mesgarnejad, A. Karma, Phase field modeling of chemomechanical fracture of intercalation electrodes: Role of charging rate and dimensionality, J. Mech. Phys. Solids. 132 (2019) 103696. https://doi.org/10.1016/j.jmps.2019.103696.

[40]  B.X. Xu, Y. Zhao, P. Stein, Phase field modeling of electrochemically induced fracture in Li-ion battery with large deformation and phase segregation, GAMM Mitteilungen. 39 (2016) 92–110. https://doi.org/10.1002/gamm.201610006.

[41]  M. Ahmadi, A hybrid phase field model for fracture induced by lithium diffusion in electrode particles of Li-ion batteries, Comput. Mater. Sci. 184 (2020) 109879. https://doi.org/10.1016/j.commatsci.2020.109879.

[42]  M. Klinsmann, D. Rosato, M. Kamlah, R.M. McMeeking, Modeling Crack Growth during Li Extraction in Storage Particles Using a Fracture Phase Field Approach, J. Electrochem. Soc. 163 (2016) A102–A118. https://doi.org/10.1149/2.0281602jes.

[43]  B. Liu, J. Xu, Cracks of silicon nanoparticles in anodes: Mechanics– Electrochemical-coupled modeling framework based on the phase-field method, ACS Appl. Energy Mater. 3 (2020) 10931–10939. https://doi.org/10.1021/acsaem.0c01916.

[44]  J.M. Allen, P.J. Weddle, A. Verma, A. Mallarapu, F. Usseglio-Viretta, D.P. Finegan, A.M. Colclasure, W. Mai, V. Schmidt, O. Furat, D. Diercks, T. Tanim, K. Smith, Quantifying the influence of charge rate and cathode-particle architectures on degradation of Li-ion cells through 3D continuum-level damage models, J. Power Sources. 512 (2021) 230415. https://doi.org/10.1016/j.jpowsour.2021.230415.

[45]  K. A. Thomas.-Alyea., J. Newman, Electrochemical Systems, 3rd ed., Wiley, 2004.

[46]  G.A. Francfort, J.J. Marigo, Revisiting brittle fracture as an energy minimization

problem, J. Mech. Phys. Solids. 46 (1998) 1319–1342. https://doi.org/10.1016/S0022-5096(98)00034-9.

[47] G.A. Francfort, B. Bourdin, J.J. Marigo, The variational approach to fracture, J. Elasticity. 91 (2008) 5-148. https://doi.org/10.1007/s10659-007-9107-3.

[48] E. Tanné, T. Li, B. Bourdin, J.J. Marigo, C. Maurini, Crack nucleation in variational phase-field models of brittle fracture, J. Mech. Phys. Solids. 110 (2018) 80–99. https://doi.org/10.1016/j.jmps.2017.09.006.

[49] S.R. Daemi, C. Tan, T. Volkenandt, S.J. Cooper, A. Palacios-Padros, J. Cookson, D.J.L. Brett, P.R. Shearing, Visualizing the Carbon Binder Phase of Battery Electrodes in Three Dimensions, ACS Appl. Energy Mater. 1 (2018) 3702–3710. https://doi.org/10.1021/acsaem.8b00501.

[50] C.D. Quilty, D.C. Bock, S. Yan, K.J. Takeuchi, E.S. Takeuchi, A.C. Marschilok, Probing Sources of Capacity Fade in LiNi$_{0.6}$Mn$_{0.2}$Co$_{0.2}$O$_2$ (NMC622): An Operando XRD Study of Li/NMC622 Batteries during Extended Cycling, J. Phys. Chem. C. 124 (2020) 8119–8128. https://doi.org/10.1021/acs.jpcc.0c00262.

[51] P. Zhu, H.J. Seifert, W. Pfleging, The ultrafast laser ablation of Li(Ni0.6Mn0.2Co0.2)O2 electrodes with high mass loading, Appl. Sci. 9 (2019) 4067. https://doi.org/10.3390/app9194067.

[52] J.J. Bailey, T.M.M. Heenan, D.P. Finegan, X. Lu, S.R. Daemi, F. Iacoviello, N.R. Backeberg, O.O. Taiwo, D.J.L. Brett, A. Atkinson, P.R. Shearing, Laser-preparation of geometrically optimised samples for X-ray nano-CT, J. Microsc. 267 (2017) 384–396. https://doi.org/10.1111/jmi.12577.

[53] S. Berg, D. Kutra, T. Kroeger, C.N. Straehle, B.X. Kausler, C. Haubold, M. Schiegg, J.

Ales, T. Beier, M. Rudy, K. Eren, J.I. Cervantes, B. Xu, F. Beuttenmueller, A. Wolny, C. Zhang, U. Koethe, F.A. Hamprecht, A. Kreshuk, Ilastik: Interactive Machine Learning for (Bio)Image Analysis, Nat. Methods. 16 (2019) 1226–1232. https://doi.org/10.1038/s41592-019-0582-9.

[54] C. Miehe, M. Hofacker, F. Welschinger, A phase field model for rate-independent crack propagation: Robust algorithmic implementation based on operator splits, Comput. Methods Appl. Mech. Eng. 199 (2010) 2765–2778. https://doi.org/10.1016/j.cma.2010.04.011.

[55] H.J. Noh, S. Youn, C.S. Yoon, Y.K. Sun, Comparison of the structural and electrochemical properties of layered Li[Ni$_x$Co$_y$Mn$_z$]O2 (x = 1/3, 0.5, 0.6, 0.7, 0.8 and 0.85) cathode material for lithium-ion batteries, J. Power Sources. 233 (2013) 121–130. https://doi.org/10.1016/j.jpowsour.2013.01.063.

[56] A.M. Grillet, T. Humplik, E.K. Stirrup, S.A. Roberts, D.A. Barringer, C.M. Snyder, M.R. Janvrin, C.A. Apblett, Conductivity Degradation of Polyvinylidene Fluoride Composite Binder during Cycling: Measurements and Simulations for Lithium-Ion Batteries, J. Electrochem. Soc. 163 (2016) A1859–A1871. https://doi.org/10.1149/2.0341609jes.

[57] L.O. Valøen, J.N. Reimers, Transport Properties of LiPF$_6$-Based Li-Ion Battery Electrolytes, J. Electrochem. Soc. 152 (2005) A882-A891. https://doi.org/10.1149/1.1872737.

[58] R. Koerver, W. Zhang, L. De Biasi, S. Schweidler, A.O. Kondrakov, S. Kolling, T. Brezesinski, P. Hartmann, W.G. Zeier, J. Janek, Chemo-mechanical expansion of lithium electrode materials-on the route to mechanically optimized all-solid-state batteries, Energy Environ. Sci. 11 (2018) 2142–2158.

https://doi.org/10.1039/c8ee00907d.

[59] G. Liu, H. Zheng, S. Kim, Y. Deng, A.M. Minor, X. Song, V.S. Battaglia, Effects of Various Conductive Additive and Polymeric Binder Contents on the Performance of a Lithium-Ion Composite Cathode, J. Electrochem. Soc. 155 (2008) A887-A892. https://doi.org/10.1149/1.2976031.

[60] T.M.M. Heenan, A. Wade, C. Tan, J.E. Parker, D. Matras, A.S. Leach, J.B. Robinson, A. Llewellyn, A. Dimitrijevic, R. Jervis, P.D. Quinn, D.J.L. Brett, P.R. Shearing, Identifying the Origins of Microstructural Defects Such as Cracking within Ni-Rich NMC811 Cathode Particles for Lithium-Ion Batteries, Adv. Energy Mater. 2002655 (2020). https://doi.org/10.1002/aenm.202002655.

[61] S.H. Park, R. Tian, J. Coelho, V. Nicolosi, J.N. Coleman, Quantifying the Trade-Off between Absolute Capacity and Rate Performance in Battery Electrodes, Adv. Energy Mater. 9 (2019) 1901359. https://doi.org/10.1002/aenm.201901359.

[62] M. Xu, B. Reichman, X. Wang, Modeling the effect of electrode thickness on the performance of lithium-ion batteries with experimental validation, Energy. 186 (2019) 115864. https://doi.org/10.1016/j.energy.2019.115864.

[63] K. Zhao, M. Pharr, J.J. Vlassak, Z. Suo, Fracture of electrodes in lithium-ion batteries caused by fast charging, J. Appl. Phys. 108 (2010) 073517. https://doi.org/10.1063/1.3492617.

[64] V.J. Ovejas, A. Cuadras, Effects of cycling on lithium-ion battery hysteresis and overvoltage, Sci. Rep. 9 (2019) 14875. https://doi.org/10.1038/s41598-019-51474-5.